\documentstyle[aps,prl]{revtex}

\def\l{\hspace*{0.05cm}}
\def\esp{\hspace*{1cm}}

\def\be{\begin{equation}}
\def\ee{\end{equation}}
\def\bea{\begin{eqnarray}}
\def\eea{\end{eqnarray}}

\begin{document}

\title{Annihilation amplitudes and factorization in
$B^\pm\to\phi K^{\ast\pm}$}

\author{L.N. Epele, D. G\'omez Dumm, A. Szynkman}

\address{IFLP, Depto. de F\'{\i}sica, Universidad Nacional de La Plata,
C.C. 67, 1900 La Plata, Argentina}

\maketitle

\begin{abstract}
We study the decay $B^\pm\to \phi K^{\ast\pm}$, followed by the decay of the
outgoing vector mesons into two pseudoscalars. The analysis of angular
distributions of the decay products is shown to provide useful
information about the annihilation contributions and possible tests of
factorization.
\end{abstract}

\vspace{1cm}


\vspace{1cm}

\section{Introduction}

The analysis of $B$ meson physics offers an attractive opportunity to
get a deep insight into the flavor structure of the Standard Model (SM)
and the origin of CP violation. In view of the wide variety of decay
channels, one can look for many different observables, providing
stringent test for the consistency of the model. However, the potential
power of the analysis is severely limited by our present theoretical
capability of dealing with strong interactions in the intermediate and
low energy regimes. In fact, only a limited number of observables are
free of theoretical uncertainties within the SM. The main sources of
theoretical errors arise from the evaluation of weak transition
amplitudes (matrix elements of quark current-current operators between
hadron states) and from the estimation of final state interaction (FSI)
effects. Thus, the theoretical control of these uncertainties turns out
to be a crucial goal.

For nonleptonic $B$ meson decays, the usual procedure to calculate the
weak transition amplitudes is based on the effective Hamiltonian
approach and the use of Wilson operator product expansion. The Wilson
coefficients contain the information from short-distance physics and
can be computed perturbatively. This program has been fully carried out
already up to next-to-leading order~\cite{Buc96}, and the main
theoretical problem to be addressed in this sense is the analysis of
long-distance physics, i.e., the computation of matrix elements of the
effective four-quark operators between hadron states. To deal with
this, a simple and widely used approach is the so-called factorization
approximation (FA) \cite{Bau87}. The extent of the validity of this
approximation is however controversial. In the last years, new
approaches, such as the so-called QCD factorization (QCDF) \cite{Ben99}
and perturbative QCD (PQCD) \cite{Keu00} schemes, have been proposed
with the aim of improving the factorization assumption on QCD grounds
\cite{Bur00}.

In the framework of FA, an extensive analysis of the phenomenology of
$B$ decays has been presented by Ali, Kramer and L\"u~\cite{Ali98},
where the authors calculate the branching fractions for charmless
non-leptonic two-body $B$ decays and propose a number of tests for the
approach. In particular, the authors in \cite{Ali98} take into account
the effects of annihilation amplitudes, which are neglected by a priori
arguments in most works on the subject. In contrast to the general
custom, it is pointed out that the contribution of annihilation
diagrams could play a significant and even dominant role, especially
in some cases where the non-annihilation amplitudes are suppressed. It
is worth to notice that the theoretical control of annihilation
amplitudes is very important for the analysis of CP-violating
observables, since in many cases the annihilation contribution carries
a weak phase different from that provided by the tree or penguin
amplitudes. This is e.g.\ the case of the decays $B^+\to K^+
\pi^0,\;\pi^+ K^0$, which have been largely analyzed in connection with
the experimental determination of the weak phase angle
$\gamma$~\cite{Neu98}. Moreover, even if in most cases annihilation
amplitudes appear to be Cabibbo-suppressed, their presence can be
important since they can compete with possible manifestations
of new physics, which could be revealed through the analysis of
CP-violating observables. On the other hand, the measurement of
annihilation contributions is interesting by itself from the point of
view of the understanding of low energy dynamics and the viability of
the theoretical approaches. For example, annihilation amplitudes are
assumed to be suppressed by powers of $\Lambda_{QCD}/m_b$ in the
framework of QCDF, while this is not the case in PQCD.

In this paper, we focus our attention in the annihilation contributions
to the process $B\to\phi K^\ast$, which is the first observed
\cite{CLE01} charmless $B$ decay into two vector mesons and has been
recently analyzed within both QCDF \cite{Cheng01} and PQCD
\cite{Chen02}. While annihilation contributions are expected to be
highly suppressed in the case of $B\to PP$ decays, an equivalent
suppression mechanism is not obvious for $B \to PV$ and $B \to VV$
processes \cite{Ali98}. For example, in the case of the decay $B^+ \to
K^{\ast +} \bar K^0$, it has been noticed that once the annihilation
part of the amplitude is taken into account, the branching ratio could
reach ---under reasonable assumptions on form factors--- an order of
magnitude higher than the value obtanied from the penguin contribution
alone \cite{Ali98}. Owing to the large theoretical uncertainties,
however, the role of annihilation contributions is in general quite
difficult to estimate from the sole measurement of branching ratios. In
this sense, $B$ decays into two vector mesons (which subsequently decay
into two particles each) present an important feature: the analysis of
angular distributions of the final outgoing particles allows to measure
both total decay rates and strong and weak phases of the contributing
amplitudes. This can be exploited e.g.\ to get different observables
for CP violating parameters \cite{Kra92,Kra93,Lon00} and solve the
so-called discrete ambiguities \cite{Cheng02}, or to analyze the
significance of the contribution of electroweak penguins \cite{Atw98}.
We show here that, in the framework of the Standard Model, the analysis
of angular distributions in the decay $B^\pm\to\phi K^{\ast\pm}$ can be
used to estimate the annihilation contributions to the process and to
test the viability of the factorization assumptions. The process
$B^\pm\to\phi K^{\ast\pm}$ is particularly interesting, since on one
hand it is expected to be dominated by penguin-like contributions
---thus annihilation amplitudes could be relatively significant--- and
on the other hand penguin and annihilation contributions carry
different weak phases, hence they can be disentangled by looking at
CP-odd terms in the angular distribution of final states.

Section II includes a general description of angular distributions and
observables in $B\to VV$ decays, while in Sect.\ III we analyze the
particular case of $B^\pm\to \phi K^{\ast\pm}$. The expected results
within the factorization approach are discussed in Sect.\ IV, and in
Sect.\ V we present some concluding remarks.

\section{Observables and angular distributions in $B\to VV$}

Let us consider the decay of a $B$ meson into two vector mesons, $B\to
V_1 V_2$, followed by the decay of both $V_1$ and $V_2$ into two
pseudoscalars $P_1 P'_1$ and $P_2 P'_2$ respectively. Following the
notation in Ref.~\cite{Wol00}, the normalized differential angular
distribution can be written as
\begin{eqnarray}
\frac{1}{\Gamma_0} \frac{d^3 \Gamma}{d \cos \theta_1 \l d \cos \theta_2
\l d \psi} & = & \frac{9}{8 \pi {\cal K}} \;\Bigg\{ K_1 \l \cos^2
\theta_1 \l \cos^2 \theta_2 \l + \l \frac{K_2}{2} \l \sin^2 \theta_1 \l
\sin^2 \theta_2 \l \cos^2 \psi \nonumber \\
& & + \l \frac{K_3}{2} \l
\sin^2 \theta_1 \l \sin^2 \theta_2 \l \sin^2 \psi \l + \l \frac{K_4}{2
\sqrt{2}} \l \sin 2 \theta_1 \l \sin 2 \theta_2 \l \cos \psi \nonumber \\
& & - \l \frac{K_5}{2 \sqrt{2}} \l \sin 2 \theta_1 \l \sin 2 \theta_2
\l \sin \psi \l - \l \frac{K_6}{2} \l \sin^2 \theta_1 \l \sin^2 \theta_2
\l \sin 2 \psi \Bigg\}\,,
\label{dg}
\end{eqnarray}
where $\theta_1$ ($\theta_2$) is the angle between the three-momentum of
$P_1$ ($P_2$) in the $V_1$ ($V_2$) rest frame and the three-momentum of
$V_1$ ($V_2$) in the $B$ rest frame, and $\psi$ is the angle between the
planes defined by the $P_1 P'_1$ and $P_2 P'_2$ three-momenta in the $B$
rest frame. The coefficients $K_i$ can be written in terms of three
independent amplitudes, $A_0$, $A_\|$ and $A_\bot$, which correspond to
the different polarization states of the vector mesons $V_1$ and
$V_2$~\cite{Dig96}. One has
\begin{eqnarray}
K_1 = |A_0|^2\;, \hspace{2.cm} & K_4 = {\rm Re} [A_{\|} A^*_0]\;, \nonumber
\\ & & \nonumber \\ K_2 = {|A_{\|}|}^2\;, \hspace{2.cm} & K_5 = {\rm Im}
[A_\bot A^*_0]\;, \nonumber \\ & & \nonumber \\ K_3 = |A_{\bot}|^2\;,
\hspace{2.cm} & K_6 = {\rm Im} [A_{\bot} A^*_{\|}]\;, \label{K}
\end{eqnarray}
and ${\cal K}\equiv K_1+K_2+K_3$. Notice that only six from the nine
possible observables given by the squared amplitude $A^\ast A$ can be
measured independently. This is due to the fact that both $V$ mesons
are assumed to decay into two spin zero particles.

In the literature, $B\to VV$ decays are also frequently described using
the helicity basis. According to their Lorentz structure, the amplitudes
can be parameterized in general as~\cite{Kra92}
\begin{eqnarray}
H_{\lambda} = \varepsilon_{1 \mu}^* (\lambda) \l \varepsilon_{2
\nu}^* (\lambda) \left [ a g^{\mu \nu} + \frac{b}{m_1 m_2} p^{\mu}
p^{\nu} + \frac{i c}{m_1 m_2} \epsilon^{\mu \nu \alpha \beta} p_{1
\alpha} p_{\beta} \right ]\;,
\label{hlam}
\end{eqnarray}
where $p$ is the $B$ meson momentum, $\lambda$ is the helicity of both
vector mesons, and $m_i$, $p_i$ and $\varepsilon_i$ stand for their
masses, momenta and polarization vectors respectively. In this way, for
$\lambda=0,\pm 1$ the helicity amplitudes are given by
\begin{equation}
H_{\pm 1} = a \pm c \l \sqrt{x^2 - 1}\;,
\esp
H_0 = - a x - b \l (x^2 - 1)\;,
\label{a}
\end{equation}
where $x \equiv (m_B^2 - m^2_1 - m^2_2)/(2 m_1 m_2)$. The relation between the
amplitudes in both schemes is
\begin{eqnarray}
A_{\bot} \l = \l \frac{H_{+1} - H_{-1}}{\sqrt{2}}, \esp
A_{\|} \l = \l \frac{H_{+1} + H_{-1}}{\sqrt{2}}, \esp A_0 \l =
\l H_0 \label{cb}
\end{eqnarray}
and the coefficients $K_i$ can be written in terms of the parameters $a$,
$b$, $c$ as
\begin{eqnarray}
\lefteqn{K_1 = |x\, a + (x^2-1)\, b|^2} \hspace*{5.5cm} & &
K_4 = -\sqrt{2}\,\left[ x\, |a|^2 + (x^2-1)\,{\rm Re}(a^\ast b)\right]
\nonumber \\
\lefteqn{K_2 = 2\,|a|^2} \hspace*{5.5cm} & &
K_5 = \sqrt{2\,(x^2-1)} \left[ x\, {\rm Im} (a c^\ast)
+ (x^2-1)\,{\rm Im}(b c^\ast)\right] \nonumber \\
\lefteqn{K_3 = 2\,(x^2-1)|c|^2} \hspace*{5.5cm} & &
K_6 = 2\sqrt{x^2-1}\; {\rm Im} (c a^\ast)
\end{eqnarray}
Relative decay rates into $V$ meson states with longitudinal and transverse
polarizations are thus given by
\begin{eqnarray}
\frac{\Gamma_L}{\Gamma_0} &  = &
\frac{|H_0|^2}{|H_0|^2+|H_{+1}|^2+|H_{-1}|^2}
= \frac{K_1}{\cal K}\;\; , \nonumber \\
\frac{\Gamma_T}{\Gamma_0} &  = &
\frac{|H_{+1}|^2+|H_{-1}|^2}{|H_0|^2+|H_{+1}|^2+|H_{-1}|^2}
= \frac{K_2+K_3}{\cal K}\;\; .
\end{eqnarray}

In general, the parameters $a$, $b$ and $c$ are complex numbers. If it
is assumed that the total decay amplitude arises as the sum of several
interfering contributions (e.g.\ different isospin channels), one has
\begin{equation}
a = \sum_i |a_i| \; e^{i \, (\delta^a_i + \varphi^a_i)}\;,
\label{contrib}
\end{equation}
where $\delta$ and $\varphi$ stand for ``strong'' (CP-conserving) and
``weak'' (CP-violating) phases respectively. Within the Standard Model,
the latter arise from the CKM matrix coefficients entering the
amplitude, while strong phases receive both contributions from short-
and long-distance physics. Similar relations as that in (\ref{contrib})
can be written for parameters $b$ and $c$.

In our analysis we will take into account both the decay $B^+\to \phi
K^{\ast +}$ and its CP-conjugated process, $B^-\to\phi K^{\ast -}$.
Following standard notation, CP-conjugated amplitudes are denoted as
$\bar A_\eta$ and $\bar H_\lambda$, with $\eta=0,\|,\bot$ and $\lambda =
0,\pm 1$. Accordingly, in the differential decay amplitude (\ref{dg}),
one should replace $K_i\to \bar K_i$ for $i=1,\dots 4$ and
$K_i\to -\bar K_i$ for $i=5,6$, which corresponds to replace $a\to\bar
a$, $b\to \bar b$ and $c\to -\bar c$ in (\ref{hlam}). Since only weak
phases change sign after a CP conjugation, one has
\begin{equation}
\bar a = \sum_i |a_i| \; e^{i \, (\delta^a_i - \varphi^a_i)}\;,
\end{equation}
while similar relations hold for $\bar b$ and $\bar c$.

\section{Penguin and annihilation amplitudes in $B^\pm\to\phi K^{\ast\pm}$}

Let us now focus on the decay $B^-\to\phi K^{\ast -}$. In the Standard
Model, this process is driven by both penguin and annihilation
contributions, with the salient feature that they carry different
weak phases. Up to small ${\cal O}(\lambda^2)$ corrections
($\lambda=|V_{ud}|\simeq 0.22$), the penguin amplitude is proportional
to the $V_{CKM}$ elements $V_{tb} V_{ts}^\ast$, while the annihilation
contribution carries the factor $V_{ub} V_{us}^\ast$. The relative phase
between both terms, up to ${\cal O}(\lambda^2)$ corrections, is
\begin{equation}
\arg \left( \frac{V_{ub} V_{us}^\ast}{V_{tb} V_{ts}^\ast} \right) \simeq
\arg \left( -\frac{V_{ud} V_{ub}^\ast}{V_{cd} V_{cb}^\ast} \right)
\equiv \gamma
\end{equation}
which is one of the angles of the so-called {\em unitarity triangle}.
Though the annihilation contribution is doubly Cabibbo suppressed with
respect to the penguin one, this is compensated by the relation between
the corresponding Wilson coefficients. We come back to this in the next
Section.

As stated in the Introduction, there is a strong theoretical motivation
to know the magnitude of annihilation amplitudes. While the penguin
contributions can be (at least, roughly) estimated with the aid of the
factorization approach, the annihilation contributions in $B\to VV$
decays are much more uncertain, since the corresponding form factors
cannot be related to semileptonic decay amplitudes. Since there are no
tree amplitudes contributing to $B^\pm\to\phi K^{\ast\pm}$, this
process is a promising one, in the sense that penguin and annihilation
contributions can be comparable in size \cite{Ali98} and their relative
magnitude can be measured. Moreover, in view of the different weak
phase structure, both contributions can be disentangled by looking at
CP-odd observables.

According to the general analysis in Section II, the coefficients $a$,
$b$ and $c$ for the case of $B^-\to\phi K^{\ast -}$ can be written as
\begin{eqnarray}
a & = & ( a_P \, e^{i\delta'_a} + a_A \, e^{i\gamma})\, e^{i\delta_a}
\nonumber \\
b & = & ( b_P \, e^{i\delta'_b} + b_A \, e^{i\gamma})\, e^{i\delta_b}
\nonumber \\
c & = & ( c_P \, e^{i\delta'_c} + c_A \, e^{i\gamma})\, e^{i\delta_c}
\label{abc}
\end{eqnarray}
where the subindices $P$ and $A$ correspond to penguin and annihilation
contributions respectively. Without loss of generality, strong phases
accompanying both terms have been separated into a global phase
$\delta_i$ and a relative phase $\delta'_i$, while $a_{P,A}$, $b_{P,A}$
and $c_{P,A}$ are real numbers. For the CP-conjugated decay
$B^+\to\phi K^{\ast +}$ the corresponding coefficients $\bar a$, $\bar
b$ and $\bar c$ are similar to those in (\ref{abc}), just changing
$\gamma\to -\gamma$.

Now, in principle, from the angular analysis of $B^\pm\to\phi
K^{\ast\pm}$ decays one can measure 12 observables, $K_i$ and $\bar
K_i$ with $i=1$ to 6. Let us first concentrate in the observables given
by the transverse modes of the vector mesons $\phi$ and $K^\ast$, that
means $i=2$, 3 and 6. With the above definitions, one has
\begin{eqnarray}
K_2 & = & 2 \left[ a_P^2 + a_A^2 + 2\, a_P\, a_A
\cos(\delta'_a-\gamma)\right] \nonumber \\
K_3 & = & 2\, (x^2-1) \left[ c_P^2 + c_A^2 + 2\, c_P\, c_A
\cos(\delta'_c-\gamma)\right]
\end{eqnarray}
and similar relations hold for $\bar K_2$ and $\bar K_3$, changing the
sign in front of $\gamma$. The relative magnitude of the annihilation
contributions can be measured from the combined observables
\begin{eqnarray}
K_2 - \bar K_2 & = & 8 \, a_P\, a_A \sin\delta'_a \sin\gamma
\nonumber \\
K_3 - \bar K_3 & = & 8 \, (x^2-1) \, c_P\, c_A \sin\delta'_c \sin\gamma
\label{k2k3}
\end{eqnarray}
which are odd under CP. A significant asymmetry provided by any of the
quantities in (\ref{k2k3}) would signal the presence of an important
annihilation contribution. This would be e.g.\ in agreement with the
prediction given by PQCD, where annihilation amplitudes are found to
enhance the decay width $\Gamma_T$ by about a factor 2 \cite{Chen02}.

Notice that, in order to be different from zero, the quantities defined
in (\ref{k2k3}) require the presence of nonzero relative strong phases
$\delta'_{a,c}$. The latter are expected to be nonvanishing even in the
absence of final state interaction effects, since in general the
penguin amplitudes include absorptive contributions \cite{Ban79}.
However, it is possible that these absorptive parts turn out to be
suppressed, hence the asymmetries in (\ref{k2k3}) could be too small to
be observed experimentally. This happens e.g.\ in the framework of
factorization, where absorptive contributions entering the effective
Wilson coefficients appear to be $\alt 20\%$ of the dispersive parts
\cite{Ali98,Ali98-2}. If this is the case, the significance of
annihilation contributions can be still estimated by considering the
observables $K_6$ and $\bar K_6$, which arise from the interference
between the amplitudes $A_\|$ and $A_\bot$. In general, the CP-odd
observable $K_6-\bar K_6$ is given by
\begin{equation}
K_6 - \bar K_6 = 4\,\sqrt{x^2-1}\,
\left[\, a_P\, c_A\, \cos (\delta_c-\delta_a-\delta'_a) -
a_A\, c_P\, \cos (\delta_c-\delta_a+\delta'_c)\right]\sin\gamma \;,
\label{k6}
\end{equation}
which is still nonzero in the limit of vanishing strong phases.
Moreover, in that case both $K_6$ and $\bar K_6$ provide separate
measurements of CP violation, obeying
\begin{equation}
K_6 = -\bar K_6 = 2\,\sqrt{x^2-1}\,
\left[\, a_P\, c_A  - a_A\, c_P\right]\sin\gamma \;.
\end{equation}
The validity of this relation would imply the presence of a significant
annihilation contribution and support the assumption that strong phases
are negligibly small.

The remaining observables $K_i$ and $\bar K_i$ with $i=1$, 4 and 5 can
also be analyzed, and once again the measurement of any significant
asymmetry $K_i - \bar K_i$ would signal the presence of annihilation
contributions within the SM. Here we do not enter in the detailed
analysis of these observables since the expressions in terms of Lorentz
invariant parameters $a$, $b$ and $c$, as well as the theoretical
analysis of form factors, turn out to be more complicated and do not
provide new physical insights.

\section{Factorization}

In the framework of factorization, the measurement of the observables
$K_i$ and $\bar K_i$ in the decay $B^\pm\to\phi K^{\ast\pm}$ can be
used not only to estimate the values of form factors related with
annihilation amplitudes, but also to test the consistency of the
approach itself. As before, we concentrate here in the observables
related to the transverse modes of the $\phi$ and $K^\ast$, that means
to $i=2$, 3 and 6.

The penguin amplitudes can be computed within generalized factorization
making use of the effective Hamiltonian approach. Once the matrix
elements of four-quark operators are factorized, the amplitudes can be
written in general in terms of form factors $f_V$, $V^{B\to V}(q^2)$,
$A_i^{B\to V}(q^2)$, $i = 0,1,2$, as
\begin{eqnarray}
\langle V (\varepsilon,p') | V_\mu | 0 \rangle & = & f_V \, m_V \,
\varepsilon^\ast_\mu \nonumber \\
\langle V (\varepsilon, p') | V_\mu | B (p) \rangle & = & -\,\frac{2}{m_V +
m_B} \; \epsilon_{\mu\nu\alpha\beta} \, \varepsilon^{\ast\,\nu} \,
p^\alpha {p'}^\beta \, V^{B\to V}(q^2) \nonumber \\
\langle V (\varepsilon, p') | A_\mu | B (p) \rangle & = & i\,
\frac{2\, m_V (\varepsilon^\ast\cdot q)}{q^2}\;
q_\mu\; A_0^{B\to V}(q^2) +
i\, (m_V + m_B) \left[\varepsilon^\ast_\mu - \frac{(\varepsilon^\ast
\cdot q)}{q^2}\; q_\mu \right] A_1^{B\to V} (q^2) \nonumber \\
& & -\, i\, \left[(p + p')_\mu\, - \frac{(m_B^2-m_V^2)}{q^2}
\;q_\mu\right] \frac{(\varepsilon^\ast\cdot q)}{m_V + m_B} \;
A_2^{B\to V}(q^2)\;,
\end{eqnarray}
Here $V(\varepsilon,p')$ stands for the outgoing vector meson $\phi$ or
$K^\ast$, $V_\mu$ and $A_\mu$ are the corresponding vector and
axial-vector quark currents and $q=p-p'$ is the momentum transfer. The
vector and axial-vector form factors can be estimated from the analysis
of semileptonic $B$ decays, using the ansatz of pole dominance to
account for the momentum dependences in the region of interest.

In this way the penguin amplitudes $a_P$, $b_P$ and $c_P$ read
\begin{eqnarray}
a_P & = & - |C_{eff}^{(P)}|\, m_\phi \, ( m_B + m_{K^\ast} )\, f_\phi \,
A_1^{B\to K^\ast}(m_\phi^2)
\nonumber \\
b_P & = & |C_{eff}^{(P)}|\, m_\phi \, \left(\frac{2\, m_{K^\ast}\, m_\phi}
{m_B + m_{K^\ast}} \right) f_\phi \, A_2^{B\to K^\ast}(m_\phi^2) \nonumber \\
c_P & = & |C_{eff}^{(P)}|\, m_\phi \, \left(\frac{2\, m_{K^\ast}\, m_\phi}
{m_B + m_{K^\ast}} \right) f_\phi \, V^{B\to K^\ast}(m_\phi^2)
\label{peng}
\end{eqnarray}
where
\begin{equation}
C_{eff}^{(P)} = \frac{G_F}{\sqrt{2}} \; V^\ast_{ts}\, V_{tb} \left[a_3 + a_4
+ a_5 - \frac{1}{2} ( a_7 + a_9 + a_{10} )\right]\,.
\label{ceffp}
\end{equation}
The coefficients $a_i$ can be calculated by means of renormalization
group analysis \cite{Buc96}, taking into account the experimental
values of the running coupling constants in the SM and the parameters
entering the $V_{CKM}$ matrix. They are complex numbers that include
absorptive contributions from QCD and electromagnetic penguin diagrams.
In general, the theoretical results include some dependence on the
renormalization scale (fixed at some value around the $b$ quark mass),
which can be reduced through the inclusion of QCD corrections to the
quark level matrix elements before the factorization procedure
\cite{Ali98-2}. In the so-called generalized FA, the coefficients are
explicitly written as functions of the number of colors $N_C$, which is
treated as a phenomenological parameter ($N_C^{eff}$) to be adjusted
from the analysis of the full pattern of charmless two-body $B$ decays.

On the other hand, the annihilation contributions can be analyzed
within FA taking into account form factors $f_P$, $V_1^{(A)}(q^2)$,
$V_2^{(A)}(q^2)$ and $A^{(A)}(q^2)$ defined by
\begin{eqnarray}
\langle 0 \, | A_\mu |\, B (p) \rangle & = & i \, f_B\, p_\mu \nonumber \\
p^\mu\; \langle K^\ast (\varepsilon_1, p_1) \phi (\varepsilon_2, p_2)
\,| V_\mu |\, 0 \rangle & = &
\left [ (\varepsilon_1^\ast \cdot \varepsilon_2^\ast) \,
p^2 \, V_1^{(A)}(p^2) - \, (\varepsilon_2^\ast \cdot p_1) \,
(\varepsilon_1^\ast\cdot p_2) \, V_2^{(A)}(p^2) \right ] \nonumber \\
p^\mu\,\langle K^\ast (\varepsilon_1, p_1) \phi (\varepsilon_2, p_2)
\,| A_\mu |\, 0 \rangle & = & \, i\, \epsilon_{\mu \nu \alpha \beta}
\,\varepsilon_1^{\ast\mu} \, \varepsilon_2^{\ast\nu} \, p_1^\alpha
\, p_2^\beta \, A^{(A)}(p^2)
\label{ffacann}
\end{eqnarray}
where $p = p_1+p_2$ is the $B$ meson four-momentum, $p^2=m_B^2$. In
this case the magnitude of the form factors cannot be estimated from
semileptonic processes, and they are introduced as unknown parameters.
{}From Eqs.\ (\ref{ffacann}), annihilation amplitudes $a_A$, $b_A$ and
$c_A$ are given by
\begin{eqnarray}
a_A & = & - |C_{eff}^{(A)}|\, f_B\, m_B^2\, V_1^{(A)}(m_B^2)
\nonumber \\
b_A & = & |C_{eff}^{(A)}|\, f_B\, m_\phi \, m_{K^\ast}\, V_2^{(A)}(m_B^2)
\nonumber \\
c_A & = & - |C_{eff}^{(A)}|\, f_B\, m_\phi \, m_{K^\ast}\, A^{(A)}(m_B^2)
\label{annih}
\end{eqnarray}
where
\begin{equation}
C_{eff}^{(A)} = \frac{G_F}{\sqrt{2}} \; V^\ast_{us}\, V_{ub} \, a_1\;.
\end{equation}
The annihilation diagram is dominated by a tree contribution,
carrying the coefficient $a_1$, which is close to one
\cite{Ali98,Ali98-2,Che99,Cheng99}. In contrast, the $a_i$
coefficients in $C_{eff}^{(P)}$ arise from QCD and electroweak
penguin diagrams, and their order of magnitude lies between
$10^{-2}$ and $10^{-4}$ \cite{Ali98,Ali98-2,Che99}. This
suppression of penguin amplitudes is however compensated by the
ratio between $V_{CKM}$ coefficients in $C_{eff}^{(A)}$ and
$C_{eff}^{(P)}$, which is of the order of $\lambda^2\simeq 0.05$.
In addition, annihilation form factors are further suppressed due
to the large momentum transfer at $q^2=m_B^2$, where they have to
be evaluated. In view of the theoretical uncertainty on the values
of these form factors at $q^2=0$, it is not immediate to determine
if annihilation contributions are large enough to interfere with
penguin ones. This analysis has to be done within a definite model
for the underlying QCD dynamics, and can be checked through the
measurement of CP-odd observables proposed here.

Let us come back to the observables $K_i$ and $\bar K_i$. In the
spirit of FA, strong phases originated by final state interactions can
be separated from short-distance physics, therefore they should be
common to both penguin and annihilation amplitudes. The only
relative strong phase between them arises then from the absorptive
contributions in the $a_i$ coefficients, which can be estimated
perturbatively. Moreover, this phase is the same for the amplitudes
$a$, $b$ and $c$, since the combination of $a_i$ coefficients in all
cases is that in $C_{eff}^{(P)}$. In this way, within FA we have
\begin{equation}
\delta'_a=\delta'_b=\delta'_c = \arg \left[a_3 + a_4
+ a_5 - \frac{1}{2} ( a_7 + a_9 + a_{10} )\right]\equiv\delta'\,.
\end{equation}
and the CP-even and CP-odd combinations of $K_i$ and $\bar K_i$ for
$i=2,3,6$ read
\begin{mathletters}
\begin{eqnarray}
K_2 + \bar K_2 & = & 4 \, (a_P^2 + a_A^2 + 2\, a_P\, a_A
\cos\delta'\cos\gamma) \label{Ka} \\
K_2 - \bar K_2 & = & 8 \, a_P\, a_A \sin\delta'\sin\gamma \label{Kb} \\
K_3 + \bar K_3 & = & 4 \, (x^2-1)\, (c_P^2 + c_A^2 + 2\, c_p\, c_A
\cos\delta'\cos\gamma) \label{Kc} \\
K_3 - \bar K_3 & = & 8 \, (x^2-1)\,c_P\, c_A \sin\delta'\sin\gamma \label{Kd} \\
K_6 + \bar K_6 & = & 4\,\sqrt{x^2-1}\,
\bigg\{\left[ a_P \, c_P + a_A\, c_A  + \left( c_P \, a_A + a_P\, c_A \right)\,
\cos\delta' \right]\,\sin(\delta_c-\delta_a) +
\nonumber \\
& & \left( a_A\, c_P - a_P\, c_A \right) \sin\delta'
\cos (\delta_c-\delta_a) \bigg\}\, \cos\gamma \label{Ke} \\
K_6 - \bar K_6 & = & 4\,\sqrt{x^2-1}\,
\bigg[\left( a_P \, c_A + a_A\, c_P \right)\,\sin\delta'
\sin(\delta_c-\delta_a) + \nonumber \\
& & \left( a_A\, c_P - a_P\, c_A \right) \cos\delta'
\cos (\delta_c-\delta_a) \bigg] \sin\gamma \label{Kf}
\end{eqnarray}
\label{set}
\end{mathletters}

This set of equations deserves some attention. First of all, as stated
in the preceding Section, the observables in Eqs.\ (\ref{Kb}),
(\ref{Kd}) and (\ref{Kf}) are CP-odd, thus they vanish in the limit of
vanishing annihilation amplitudes. Notice that, in the framework of
factorization, the annihilation coefficients $a_A$, $c_A$ and the
strong FSI phases $\delta_a$, $\delta_c$ are the only unknown
parameters (the former, due to the uncertainty in the estimation of
form factors), whereas there is some allowed range for the values of
$\delta'$ and $\gamma$ (the latter given by experimental measurements
of CP violation in $K$ physics and the golden plate $B\to J/\Psi K_s$).
In this way, the six-equation system (\ref{set}) is overdetermined, and
the experimental information on the observables $K_i$ and $\bar K_i$
can be used both to get a measurement of the magnitude of annihilation
contributions an to test the consistency of the approach. In
particular, the expressions in Eqs.\ (\ref{Ka}) to (\ref{Kd}) do not
depend from strong phases $\delta_{a,c}$. With the measurement of this
four observables (which corresponds to the measurement of $|A_\| |$ and
$|A_\bot |$ for $B^-\to \phi K^{\ast -}$ and $B^+\to \phi K^{\ast +}$),
and getting the estimation of penguin amplitudes from Eqs.\
(\ref{peng}), it would be possible to extract the values of
annihilation coefficients $a_A$ and $c_A$ as well as the phases
$\delta'$ and $\gamma$, and to check the consistency of the values of
these phases with the theoretical and experimental bounds. Then, Eqs.\
(\ref{Ke}) and (\ref{Kf}) provide a further check of the results with
the additional possibility of getting information on the strong phase
difference $\delta_a-\delta_c$. {}From the values of the coefficients
$a_A$ and $c_A$ it is immediate to obtain estimations for the
unknown annihilation form factors $V_1^{(A)}$ and $A^{(A)}$.

Within factorization one would also expect the strong FSI phases
$\delta_a$ and $\delta_c$ to be relatively small. In this limit (or in
the case in which they are approximately equal) Eqs.\ (\ref{Ke}) and
(\ref{Kf}) reduce to
\begin{eqnarray}
K_6 + \bar K_6 & = & 4\,\sqrt{x^2-1}\,
\left( a_A\, c_P - a_P\, c_A \right) \sin\delta' \cos\gamma \nonumber \\
K_6 - \bar K_6 & = & 4\,\sqrt{x^2-1}\,
\left( a_A\, c_P - a_P\, c_A \right) \cos\delta' \sin\gamma
\label{K6}
\end{eqnarray}
and the ratio between them is given by
\begin{equation}
\frac{K_6 - \bar K_6}{K_6 + \bar K_6} =
\frac{\tan \gamma}{\tan \delta'}
\label{fraction}
\end{equation}
which does not depend on the assumptions on form factors. This
relation allows a simple test of the significance of strong FSI
phases within FA, provided that the interference between penguin
and annihilation amplitudes is strong enough to give measurable
values for the observables in (\ref{K6}).

The above equations include two approximations that are worth to be
mentioned. In fact, penguin contributions should also include the
so-called annihilation penguin diagrams, which carry the same weak
phase as in $C_{eff}^{(P)}$. It can be seen that the corresponding
combination of $a_i$ coefficients is different from that in Eq.\
(\ref{ceffp}), even if the order of magnitude is not significantly
modified \cite{Che99}. Within FA, these amplitudes involve annihilation
matrix elements, therefore their contributions to $a_P$, $b_P$ and
$c_P$ are proportional to annihilation form factors. Although the
inclusion of these terms does not introduce more unknown parameters in
Eqs.\ (\ref{set}), the disentanglement of annihilation form factors
becomes more complicated. Here the contribution of annihilation
penguins has been neglected for simplicity. However, they should be
incorporated into the set of equations (\ref{set}) if the effect of
annihilation amplitudes is found to be relatively large. A second
approximation has been done when assuming that penguin contributions
carry a global weak phase arising from the $V_{CKM}$ combination
$V_{ts}^\ast V_{tb}$. Here we have neglected the contribution of a
virtual $u$ quark in the penguin loop, which carries a factor
$V_{us}^\ast V_{ub}$ and could lead to an observable signal of CP
violation due to the presence of absorptive strong phases \cite{Ban79}.
This contribution is doubly-Cabibbo suppressed with respect to the
dominant one, and the final effect is expected to be below 1\%
\cite{Ger91}. Thus a clear evidence of the presence of annihilation
amplitudes would require a minimum signal of a few percent level.

Let us conclude this Section by presenting a brief numerical analysis
of the expected results within the framework of generalized FA.
Theoretical estimations of effective coefficients for penguin
amplitudes have been performed in previous works
\cite{Ali98,Ali98-2,Che99}, leading to the approximate values quoted in
Table I for different choices of the parameter $N_C^{eff}$. The values
for $a_P$, $b_P$ and $c_P$ in the Table have been estimated following
Ref.\ \cite{Ali98}, where the relevant form factors at $q^2=0$ are
calculated combining lattice QCD results at a high $q^2$ scale with
light-cone QCD sum rule analysis. As it can be seen, $a_P$ turns out to
be kinematically enhanced with respect to $c_P$. However, in the
expressions for the observables in Eqs.\ (\ref{set}) this enhancement
is compensated by the factors $(x^2-1)$ and $\sqrt{x^2-1}$, where
$x\simeq 14$ for the process under consideration. In Table I we have
also included the estimations for the absorptive phases $\delta'$, as
well as the results for the total branching ratio for $B^\pm\to \phi
K^{\ast\pm}$ arising from penguin contributions alone. In favor of
generalized FA, the latter appear to be in agreement with recent
experimental measurements \cite{CLE01}, which quote a decay branching
fraction of $10^{-5}$ with an error of about 50\%. Nevertheless, the
theoretical results in Table I should be taken only as estimative, and
even if the experimental error in the measurement of $BR(B^\pm\to\phi
K^{\ast\pm})$ is expected to be reduced in the future, it is unlikely
that from the sole measurement of the branching ratio one could
evaluate the interference of penguin amplitudes with other possible
contributions.

Concerning the theoretical predictions for the absorptive phase
$\delta'$, it can be seen that within the approach of generalized FA
its value lies in a range between 10 and 20 degrees. The remaining
parameter to be taken into account in Eqs.\ (\ref{set}) is the
CP-violating phase $\gamma$, which can be constrained by considering
the present measurements of $V_{CKM}$ matrix elements and the
experimental results for CP-violating observables in $K$ and $B$
physics. We quote here the recent estimation in Ref.\ \cite{Bur02},
\begin{equation}
\gamma = 63.5^\circ \pm 7.0^\circ \;.
\end{equation}
These ranges for $\delta'$ and $\gamma$ can be used to constrain
the expected result for the ratio in Eq.\ (\ref{fraction}). Notice
however that this expression holds only in the limit in which penguin
annihilation amplitudes are neglected.

Finally, since we have concentrated here in observables related to $B$ decays
into transversely polarized vector mesons $\phi$ and $K^\ast$, it is
important to notice that the values in Table I lead to a relative decay
fraction $\Gamma_T/\Gamma_0 \simeq 0.14$. Once again this value
corresponds to the penguin contribution alone, therefore it does not
depend on the global factor $|C_{eff}^{(P)}|$ which carries the
dependence on $N_C^{eff}$. If this ratio is not significantly reduced
after the inclusion of annihilation amplitudes, the analysis of
$B^\pm\to \phi K^{\ast\pm}$ decays would include enough statistics so
as to allow the measurements of the observables $K_i$ and $\bar K_i$ in
Eqs.\ (\ref{set}) in the near future.

\section{Conclusions}

We study the decay $B^\pm \to \phi K^{\ast\pm}$, showing that the
analysis of angular distributions of the final outgoing particles can
be used to estimate the significance of annihilation contributions to
the decay amplitude. The magnitude of these contributions represents
an interesting subject from the theoretical point of view, in view
e.g.\ of the different predictions obtained from QCD-based approaches
such as PQCD or QCDF.

In general, due to the existing hadronic uncertainties in the
estimation of amplitudes, annihilation contributions are quite
difficult to evaluate from the experimental information on total
branching ratios. Here we point out that the decay $B^\pm \to \phi
K^{\ast\pm}$ offers an interesting opportunity in this sense, since
annihilation amplitudes may be relatively large, and they can be
disentangled by looking at certain CP-odd observables. In particular,
in the framework of factorization, the experimental information can be
used to measure annihilation form factors and strong final state
interaction phases. The analysis also serves as a test of the
consistency of the factorization approach, taking into account the
theoretical estimation of the coefficients in the effective $\Delta
B=1$ Hamiltonian and the experimental information on the angle $\gamma$
of the unitarity triangle.

\acknowledgements

We are grateful to L.\ de Paula for providing useful information
and Y.\ Nir for revising the manuscript. D.G.D.\ acknowledges
financial aid from Fundaci\'on Antorchas (Argentina). This work has
been partially supported by CONICET and ANPCyT (Argentina).

\begin{table}
\begin{center}
\begin{tabular}{ccccccc}
$\ \ N_C^{eff} \ \ $ &  $(G_F/\sqrt{2})^{-1}|C_{eff}^{(P)}|$
& $a_P$ [GeV] & $b_P$ [GeV] & $c_P$ [GeV] &  $\delta'$ & $BR$ \\
    \hline
2 & $2.2\times 10^{-3}$ & $-0.9\times 10^{-8}$ & $4.2\times 10^{-10}$
  & $0.6\times 10^{-9}$ & $10^\circ$ & $2.0\times 10^{-5}$ \\
3 & $1.6\times 10^{-3}$ & $-0.7\times 10^{-8}$ & $3.1\times 10^{-10}$
  & $4.3\times 10^{-10}$ & $11^\circ$ & $1.1\times 10^{-5}$ \\
$\infty$ & $3.7\times 10^{-4}$ & $-1.6\times 10^{-9}$ & $0.7\times 10^{-10}$
  & $1.0\times 10^{-10}$ & $18^\circ$ & $0.6\times 10^{-6}$ \\
\hline
\end{tabular}
\caption[]{Results for penguin effective coefficients and amplitudes
within the generalized factorization approach.} \label{tab1}
\end{center}
\end{table}

\end{document}